\documentclass[aps,prd,reprint,groupedaddress,nofootinbib]{revtex4-1}

\usepackage{graphicx}
\usepackage{epsfig}
\usepackage{bm}
\usepackage{amsfonts}
\usepackage{multirow}

\begin{document}


\title{Tachyon inflation with steep potentials}


\author{K. Rezazadeh}
\email[]{rezazadeh86@gmail.com}
\affiliation{Department of Physics, University of Kurdistan, Pasdaran Street, P.O. Box 66177-15175, Sanandaj, Iran}

\author{K. Karami}
\email[]{kkarami@uok.ac.ir}
\affiliation{Department of Physics, University of Kurdistan, Pasdaran Street, P.O. Box 66177-15175, Sanandaj, Iran}

\author{S. Hashemi}
\email[]{hashemi.s273@gmail.com}
\affiliation{Department of Physics, University of Kurdistan, Pasdaran Street, P.O. Box 66177-15175, Sanandaj, Iran}


\date{\today}

\begin{abstract}

Within the framework of tachyon inflation, we consider different steep potentials and check their viability in light of the Planck 2015 data. We see that in this scenario, the inverse power-law potential $V(\phi)=V_{0}(\phi/\phi_{0})^{-n}$ with $n=2$ leads to the power-law inflation with the scale factor $a(t)\propto t^{q}$ where $q>1$, while with $n<2$, it gives rise to the intermediate inflation with the scale factor $a(t)\propto\exp\left(At^{f}\right)$ where $A>0$ and $0<f<1$. We find that, although the inverse power-law potential with $n\leq 2$ is completely ruled out by the Planck 2015 data, the result of this potential for $n>2$ can be compatible with the 95\% CL region of Planck 2015 TT, TE, EE+lowP data. We further conclude that the exponential potential $V(\phi)=V_{0}e^{-\phi/\phi_{0}}$, the inverse $\cosh$ potential $V(\phi)=V_{0}/\cosh(\phi/\phi_{0})$, and the mutated exponential potential $V(\phi)=V_{0}\left[1+(n-1)^{-(n-1)}(\phi/\phi_{0})^{n}\right]e^{-\phi/\phi_{0}}$ with $n=4$, can be consistent with the 95\% CL region of Planck 2015 TT, TE, EE+lowP data. Moreover, using the $r-n_s$ constraints on the model parameters, we also estimate the running of the scalar spectral index $dn_{s}/d\ln k$ and the local non-Gaussianity parameter $f_{{\rm NL}}^{{\rm local}}$. We find that the lower and upper bounds evaluated for these observables are compatible with the Planck 2015 results.

\end{abstract}


\maketitle

\section{Introduction}
\label{section:introduction}

The inflation theory at first was proposed to solve the problems of the hot
 big bang cosmology such as the flatness problem, the horizon problem, and the magnetic monopole problem \cite{Guth1981, Linde1982, Albrecht1982, Linde1983}. This theory assumes that a very fast accelerated expansion has occurred at the early stages of our Universe. Then, is was realized that perturbations produced during inflation can reasonably explain the large scale structure (LSS) formation in the Universe as well as the anisotropies observed in the cosmic microwave background (CMB) radiation  \cite{Mukhanov1981, Hawking1982, Starobinsky1982, Guth1982}. This fact makes it possible for us to connect the late time observations to the dynamics of the inflationary era of our Universe. The inflation theory predicts that the primordial perturbations should be adiabatic and approximately scale-invariant that these predictions are in well agreement with the current observational results. Precise observational results are obtained by the Planck satellite from measuring of the anisotropies in both the temperature and polarization of the CMB radiation \cite{Planck2015, Planck2015NG}. Using the Planck 2015 data, we can check viability of different inflation models and find valuable constraints on them.

In the inflation theory, usually a scalar field is invoked to explain the accelerated expansion of the Universe. The scalar field responsible for inflation is called ``inflaton.'' The classical dynamics of the inflaton during inflation is determined by a potential. Beside the classical dynamics, the inflaton has quantum fluctuations which are the origin of the two types perturbations including the scalar and tensor perturbations. We can see the imprint of these perturbations on the LSS formation and also on the anisotropies observed in the CMB radiation.

The standard scenario of inflation is based on a canonical scalar field in the framework of Einstein gravity. Viability of various inflationary potentials within the framework of standard inflation has been extensively investigated in the literature \cite{Martin2014a, Martin2014b, Huang2016, Okada2016}. However, the effective field theory allows the higher-order scalar kinetic terms to appear in the action in the regime of high energy physics \cite{Skiba2010, Luty2005, Kaplan2005}. Since the cosmological inflation has occurred in the regime of high energy physics, therefore we expect that the noncanonical kinetic terms play an effective role in the inflationary dynamics. Inspiring this, so far some inflationary models have been proposed which consider the noncanonical kinetic terms in the action and these models are known as the noncanonical models of inflation \cite{Armendariz-Picon1999, Garriga1999, Silverstein2004, Alishahiha2004, Chen2005a, Chen2005b, Chen2007, Tolley2010, Franche2010a, Franche2010b, Bruck2011, Li2012, Unnikrishnan2012, Unnikrishnan2013, Cespedes2015, Rezazadeh2015, Geng2015, Li2016, Nazavari2016}. One important noncanonical scalar field is tachyon which has motivations from the quantum field theory as well as the string theory \cite{Sen1998, Sen1999, Sen2002a, Sen2002b, Sen2002c, Sen2005, Gerasimov2000, Kutasov2000}. The tachyon scalar field can be used in bosonic string field theory describing the decay of an unstable D-brane \cite{Sen2002a, Sen2002b, Sen2002c, Sen2005}.

Inflation with the tachyon scalar field has been already studied in the literature \cite{Padmanabhan2002, Feinstein2002, Abramo2003, Kofman2002, Gibbons2002, Fairbairn2002, Steer2004, delCampo2009, Calcagni2005, Nozari2013, Nozari2014, Li2014, Sharif2014, Aghamohammadi2014, Barbosa-Cendejas2015, Bilic2017, Herrera2006, Xiao2011, Zhang2014, Kamali2016, Motaharfar2016}. For instance, using the cosmological perturbation theory in the framework of tachyon inflation, the primordial power spectra of the scalar and tensor perturbations have been obtained in \cite{Steer2004}. Moreover, the bispectrum of this model has been derived in \cite{Calcagni2005}, where the author has shown that the amplitude of the non-Gaussianity for the tachyon inflation is of order of the slow-roll parameters, just like the standard canonical inflation \cite{Maldacena2003, Acquaviva2003, Rigopoulos2005}.

In the present work, we focus on the tachyon inflation and our main aim is to check the validity of different steep potentials in light of the current observational results deduced from the Planck 2015 data \cite{Planck2015}. This paper is organized as follows. In Sec. \ref{section:tachyon}, we study inflation driven by a tachyon scalar field. In  Secs. \ref{section:exp}-\ref{section:mexp}, we investigate several steep potentials and examine their consistency in light of the Planck 2015 data. Finally, in Sec. \ref{section:conclusions}, we summarize our conclusions.

\section{Tachyon inflation}
\label{section:tachyon}

The action of a simple inflationary model with the tachyon scalar field can be written as follows \cite{Sen2002c, Sen2005}
\begin{equation}
\label{S}
S=\int d^{4}x\sqrt{-g}\left[\frac{1}{2}R-V(\phi)\sqrt{1+g^{\mu\nu}\partial_{\mu}\phi\partial_{\nu}\phi}\right],
\end{equation}
where $\phi$ is the tachyon scalar field which has dimensions of $1/M_P$ that ${M_P} \equiv 1/\sqrt {8\pi G}$ is the reduced Planck mass. It should be noted that throughout this paper we work in the units where ${M_P} = 1$, for the sake of convenience. In the above equation, $R$ is the Ricci scalar, and $V(\phi)$ is the inflaton potential. For the tachyon scalar field, the inflaton potential $V(\phi)$ should satisfy the following conditions \cite{Gibbons2002, Fairbairn2002, Steer2004, delCampo2009}
\begin{equation}
\label{V,tachyon}
 V(\phi  \to \infty ) \to 0, \qquad\frac{dV}{d\phi} < 0.
\end{equation}
The potentials which satisfy these conditions are so-called the ``steep potentials''.
{It should be noted that the tachyon potential has a minimum at $\phi  \to \infty$, and hence the inflaton cannot oscillate around its minimum \cite{Kofman2002}. Consequently, the conventional  reheating mechanism is not applicable in the tachyon inflationary scenario. Indeed, in the tachyon inflation, the energy density of the tachyon field always dominate that of the radiation field so that the Universe never enters the radiation-dominated era which is essential for nucleosynthesis. To overcome this problem, one can assume that the tachyon energy was fine-tuned to be exponentially small in the postinflationary Universe \cite{Kofman2002}. An alternative solution is to consider models of hybrid inflation \cite{Linde1991, Linde1994}, in which a complex field plays the role of inflaton that its potential has a minimum not at $\phi \to \infty$, but at $\left|\phi\right|\ll1$. However, hopefully future developments of the string theory may present other suggestions to resolve the reheating problem in the tachyon inflation. Also, suggestion of other mechanisms for the reheating process may resolve this drawback. For instance, the tachyon field can be considered in the warm inflationary scenario in which duo to dissipation, the inflaton gives its energy to the radiation field continuously during inflation \cite{Herrera2006, Xiao2011, Zhang2014, Kamali2016, Motaharfar2016}.}

We consider a homogeneous and isotropic Universe described by the Friedmann-Robertson-Walker (FRW) metric that we take its spatial geometry to be flat. Also, we assume that during inflation, the Universe is filled by the scalar field $\phi$ in the form of a perfect fluid with the energy-momentum tensor $T^{\mu}_{\nu}={\rm diag}(-\rho_{\phi},\,p_{\phi},\,p_{\phi},\,p_{\phi})$, where $\rho_\phi$ and $p_\phi$ denote the energy density and pressure of the scalar field, respectively. Therefore, dynamics of the Universe is determined by the Friedmann equation
\begin{equation}
\label{H,rho_phi}
H^{2}=\frac{1}{3}\rho_{\phi},
\end{equation}
together with the conservation equation
\begin{equation}
\label{dot{rho}_phi}
\dot{\rho}_{\phi}+3H\left(\rho_{\phi}+p_{\phi}\right)=0,
\end{equation}
where the dot denotes the derivative with respect to the cosmic time $t$. Also $H \equiv \dot a/a$ is the Hubble parameter, where $a$ is the scale factor of the Universe.

The energy density and pressure of the tachyon scalar field are given by \cite{Gibbons2002, Fairbairn2002, Steer2004, delCampo2009, Calcagni2005, Nozari2013, Nozari2014, Li2014, Sharif2014, Aghamohammadi2014}
\begin{eqnarray}
\label{rho_phi}
\rho_{\phi} &=& \frac{V(\phi)}{\sqrt{1-\dot{\phi}^{2}}},
\\
\label{p_phi}
p_{\phi} &=& -V(\phi)\sqrt{1-\dot{\phi}^{2}}.
\end{eqnarray}
Substituting the energy density (\ref{rho_phi}) into the conservation equation (\ref{dot{rho}_phi}) leads to the evolution equation for the tachyon scalar field as
\begin{equation}
\label{ddot{phi}}
\frac{{\ddot \phi }}{{1 - {{\dot \phi }^2}}} + 3H\dot \phi  + \frac{{V'(\phi )}}{{V(\phi )}} = 0,
\end{equation}
where the prime denotes the derivative with respect to $\phi$.

To study inflation, it is useful to define the slow-roll parameters
\begin{eqnarray}
\label{varepsilon_1}
\varepsilon_{1} &\equiv& -\frac{\dot{H}}{H^{2}},
\\
\label{varepsilon_{i+1}}
\varepsilon_{i+1} &\equiv& \frac{\dot{\varepsilon}_{i}}{H\varepsilon_{i}},\qquad(i\geq1).
\end{eqnarray}
From definition (\ref{varepsilon_1}) it is obvious that the required condition for inflation ($\ddot a > 0$) is $\varepsilon_1 < 1$.

If we assume the slow-roll conditions $\dot{\phi}^{2}\ll1$ and $\left|\ddot{\phi}\right|\ll\left|3H\dot{\phi}\right|,\,\left|V'(\phi)/V(\phi)\right|$, then the Friedmann equation (\ref{H,rho_phi}) reduces to
\begin{equation}
\label{H,V}
H^{2}\approx \frac{1}{3}V(\phi),
\end{equation}
and furthermore, the evolution equation (\ref{ddot{phi}}) for inflaton becomes
\begin{equation}
\label{dot{phi}}
3H\dot{\phi}+\frac{V'(\phi)}{V(\phi)} \approx 0.
\end{equation}

In the slow-roll regime, we can obtain the first three slow-roll parameters from Eqs. (\ref{varepsilon_1}) and (\ref{varepsilon_{i+1}}) in terms of the potential as follows
\begin{eqnarray}
\label{varepsilon_1,V}
\varepsilon_{1} &\approx& \frac{V'^{2}}{2V^{3}},
\\
\label{varepsilon_2,V}
\varepsilon_{2} &\approx& -\frac{2V''}{V^{2}}+\frac{3V'^{2}}{V^{3}},
\\
\label{varepsilon_3,V}
\varepsilon_{3} &\approx& -\frac{V'\left(9V'^{3}-10VV'V^{\prime\prime}+2V^{2}V'''\right)}{V^{3}\left(-3V'^{2}+2VV^{\prime\prime}\right)}.
\end{eqnarray}

In the study of inflation, we usually express the amount of inflation in terms of the $e$-fold number
\begin{equation}
\label{N}
N\equiv\ln\frac{a_{e}}{a},
\end{equation}
where $a_e$ refers to the scale factor at the end of inflation. The inflationary observables should be evaluated at the epoch when the perturbations exit the Hubble horizon and it is specified by the relation $k_*=a_* H_*$, where $k_*$ is the comoving wave number at the horizon exit. Throughout this paper, we consider the pivot scale $k_{*}=0.05\,{\rm Mpc}^{\rm -1}$, as adopted by the Planck 2015 collaboration \cite{Planck2015}. It has been shown that the wavelengths of the perturbations corresponding to the CMB temperature anisotropies cross the horizon around 50 to 60 $e$-folds before the end of inflation \cite{Liddle2003, Dodelson2003}. From definition (\ref{N}), we find
\begin{equation}
\label{d{N}}
dN=-Hdt=-\frac{H}{\dot{\phi}}d\phi.
\end{equation}
Using the above equation together with Eqs. (\ref{H,V}) and (\ref{dot{phi}}), we reach the following differential equation
\begin{equation}
\label{d{phi}/dN}
\frac{d\phi}{dN} \approx \frac{V'}{V^{2}}.
\end{equation}
One can solve this to find the evolution of the inflaton versus the $e$-fold number in the slow-roll approximation.

During inflation two types of perturbations can be generated, namely the scalar and tensor perturbations. The power spectrum of the scalar perturbations in the tachyon inflationary scenario and in the slow-roll approximation is given by \cite{Steer2004}
\begin{equation}
\label{mathcal{P}_s}
{\cal P}_{s} \approx \frac{1}{8\pi^{2}}\frac{H^{2}}{\varepsilon_{1}}.
\end{equation}
The reported value for the amplitude of the scalar perturbation at $k_{*}=0.05\,{\rm Mpc}^{{\rm -1}}$ is $\ln\left[10^{10}{\cal P}_{s}\left(k_{*}\right)\right]=3.094\pm0.034$, according to 68\% CL constraint from Planck 2015 TT, TE, EE+ lowP observational data \cite{Planck2015}.

The scale-dependence of the scalar power spectrum is specified by the scalar spectral index defined as
\begin{equation}
\label{n_s,definition}
n_{s}-1\equiv\frac{d\ln{\cal P}_{s}}{d\ln k}.
\end{equation}
The 68\% CL constraint of Planck 2015 TT, TE, EE+lowP data on this quantity is $n_{s}=0.9644\pm0.0049$ \cite{Planck2015}. In order to obtain an expression for $d\ln k$, we use the relation $k_*=a_* H_*$ and note that $H$ is approximately constant during slow-roll inflation. Therefore, we can obtain the relation
\begin{equation}
\label{d{ln}k}
d\ln k \approx  H dt,
\end{equation}
being valid around the horizon crossing. Using Eqs. (\ref{varepsilon_1}), (\ref{varepsilon_{i+1}}), (\ref{mathcal{P}_s}), and (\ref{d{ln}k}) in definition (\ref{n_s,definition}), we can obtain the relation of the scalar spectral index in tachyon inflation as
\begin{equation}
\label{n_s}
n_{s}=1-2\varepsilon_{1}-\varepsilon_{2}.
\end{equation}
We further can use Eqs. (\ref{varepsilon_1}), (\ref{varepsilon_{i+1}}), (\ref{d{ln}k}), and (\ref{n_s}) to obtain the running of the scalar spectral index as
\begin{equation}
\label{dn_s/d{ln}k}
\frac{dn_{s}}{d\ln k}=-2\varepsilon_{1}\varepsilon_{2}-\varepsilon_{2}\varepsilon_{3}.
\end{equation}
From this, it is clear that the value of $d{n_s}/d\ln k$ in the tachyon inflation is of the second order of the slow-roll parameters. The Planck 2015 constraint on the running of the scalar spectral index is $dn_{s}/d\ln k=-{\rm 0}.0085\pm{\rm 0}.0076$ (68\% CL, Planck 2015 TT, TE, EE+lowP) \cite{Planck2015}.

The tensor power spectrum for the tachyon inflation is same as that of the standard canonical inflation, and it is given by \cite{Steer2004}
\begin{equation}
\label{mathcal{P}_t}
{\cal P}_{t} \approx \frac{2}{\pi^{2}}H^{2}.
\end{equation}
Using the above equation together with Eq. (\ref{d{ln}k}), we can simply calculate the tensor spectral index and get
\begin{equation}
\label{n_t}
n_{t}\equiv\frac{d\ln{\cal P}_{t}}{d\ln k}=-2\varepsilon_{1}.
\end{equation}
The current experimental devices are not accurate enough to measure the tensor spectral index. However, we may be able to determine this observable by increasingly precise measurements in the future.

One other important inflationary observable, which is widely used to discriminate between inflationary models, is the tensor-to-scalar ratio
\begin{equation}
\label{r,definition}
r\equiv\frac{{\cal P}_{t}}{{\cal P}_{s}}.
\end{equation}
The upper bound from the Planck 2015 results for this observable is $r < 0.149$ (95\% CL, Planck 2015 TT, TE, EE+lowP) \cite{Planck2015}. In the framework of tachyon inflation, we can calculate it from Eqs. (\ref{mathcal{P}_s}) and (\ref{mathcal{P}_t}) as
\begin{equation}
\label{r}
r=16\varepsilon_{1}.
\end{equation}
It is easy to see that Eqs. (\ref{n_t}) and (\ref{r}) provide the consistency relation for the tachyon inflation as
\begin{equation}
\label{r,n_t}
r =  - 8{n_t},
\end{equation}
which is identical to that of the standard canonical inflation. In \cite{Steer2004, Barbosa-Cendejas2015}, the relations of the inflationary observables for the tachyon inflation have been presented to the second order of slow-roll parameters, and it has been discussed that the difference between the consistency relations of the standard canonical inflation and the tachyon inflation appears only at the quadratic order in slow-parameters.

In \cite{Calcagni2005}, the author has calculated the bispectrum of the perturbations in the tachyon inflationary scenario, and derived the so-called local non-Gaussianity parameter as\footnote{In this paper, we follow the conventional notation of the Planck Collaboration \cite{Planck2015NG} in defining the local non-Gaussianity parameter $f_{{\rm NL}}^{{\rm local}}$, and therefore Eq. (\ref{f_{NL}^{local}}) is $-5/3$ times the result of \cite{Calcagni2005}, which has adopted a different notation.}
\begin{equation}
\label{f_{NL}^{local}}
f_{{\rm NL}}^{{\rm local}}=\frac{5}{12}\left(1-n_{s}\right).
\end{equation}
It should be noted that the local non-Gaussianity parameter for the tachyon inflation is same as that of the standard canonical inflation \cite{Maldacena2003, Acquaviva2003, Rigopoulos2005}. The Planck 2015 constraints on the primordial non-Gaussianity have provided a bound on this observable as $f_{{\rm NL}}^{{\rm local}}=0.8\pm5.0$ (68\% CL, Planck 2015 T+E) \cite{Planck2015NG}.

So far, we have obtained the necessary relations for the observables in the tachyon inflationary scenario. In the subsequent sections, we will apply these relations to check viability of several steep potentials in light of the Planck 2015 data.

\section{Exponential potential}
\label{section:exp}

The first potential whose viability in the framework of tachyon inflation, we examine is the exponential potential
\begin{equation}
\label{V,exp}
V(\phi)=V_{0}e^{-\frac{\phi}{\phi_{0}}},
\end{equation}
where $V_0$ and $\phi_0$ are constant parameters. In the framework of standard canonical inflationary scenario, this potential leads to the power-law inflation with the scale factor $a(t)\propto t^{q}$ where $q>1$ \cite{Lucchin1985, Halliwell1987, Yokoyama1988, Liddle1989}, which is not compatible with the Planck 2015 data, as it has been shown in \cite{Planck2015, Rezazadeh2016, Rezazadeh2017}.

In the framework of tachyon inflation, if we use the exponential potential (\ref{V,exp}) in Eq. (\ref{d{phi}/dN}) and then solve the resulting differential equation, we find the scalar field in terms of the $e$-fold number as
\begin{equation}
\label{phi,N,exp}
\phi=-\phi_{0}\ln\left[\frac{N}{V_{0}\phi_{0}^{2}}+e^{-\frac{\phi_{e}}{\phi_{0}}}\right],
\end{equation}
where $\phi_e$ refers to the value of inflaton at the end of inflation. To determine $\phi_e$, first we simplify the first slow-roll parameter from Eq. (\ref{varepsilon_1}) and find
\begin{equation}
\label{varepsilon_1,exp}
\varepsilon_{1}=\frac{1}{2V_{0}\phi_{0}^{2}}e^{\frac{\phi}{\phi_{0}}},
\end{equation}
which is an increasing function during inflation. Therefore, using the end of inflation constraint ($\varepsilon _1=1$), we obtain
\begin{equation}
\label{phi_e,exp}
\phi_{e}=\phi_{0}\ln\left(2V_{0}\phi_{0}^{2}\right).
\end{equation}

Now, we can use Eq. (\ref{phi,N,exp}) in (\ref{mathcal{P}_s}), and obtain the scalar power spectrum as
\begin{equation}
\label{mathcal{P}_s,exp}
{\cal P}_{s}=\frac{1}{48\pi^{2}\phi_{0}^{2}}\left(2N+1\right)^{2}.
\end{equation}
One can use the above equation to fix the amplitude of the scalar perturbations at the horizon exit from the observational results, and find a constraint on the parameter $\phi_0$.

\begin{figure*}[t]
\begin{center}
\scalebox{1}[1]{\includegraphics{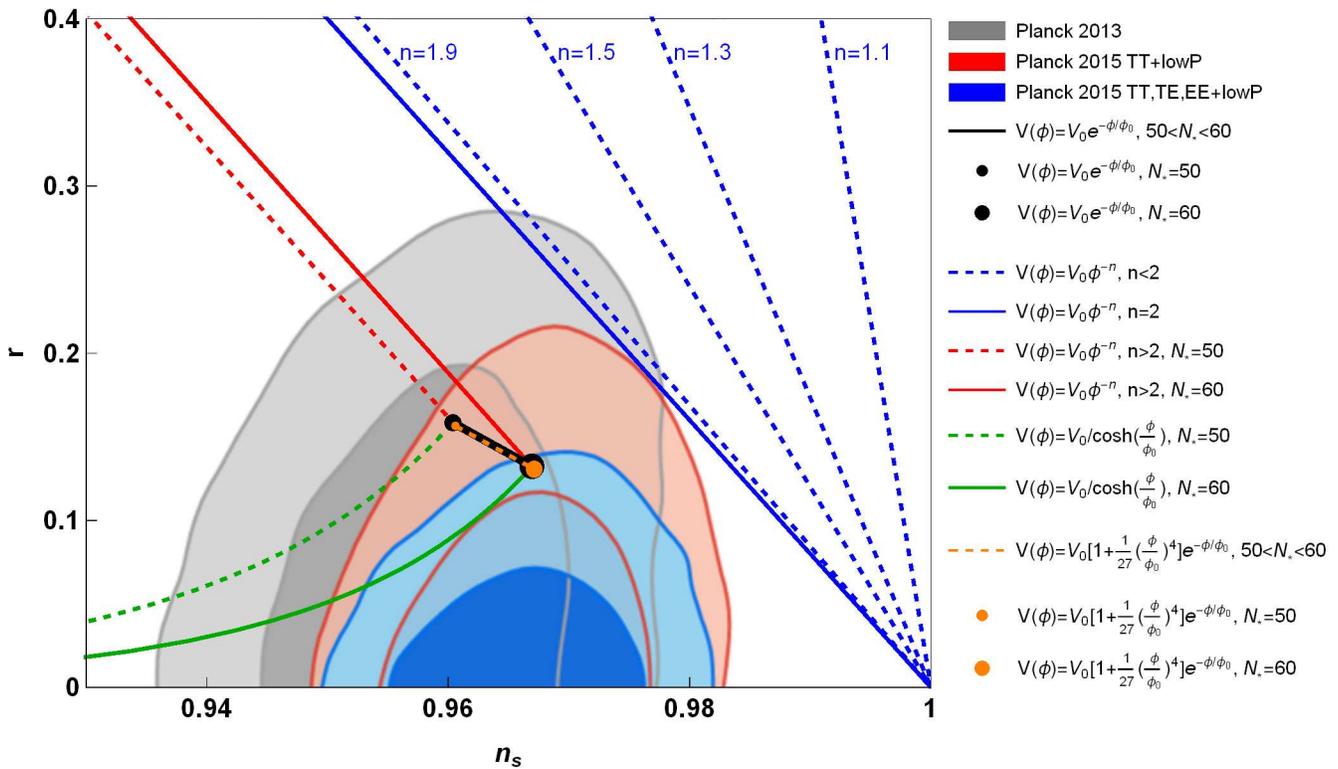}}
\caption{Predictions of different steep potentials in $r-n_s$ plane in the tachyon inflationary scenario in comparison with the observational results. The marginalized joint 68\% and 95\% CL regions from Planck 2013, Planck 2015 TT+lowP and Planck 2015 TT,TE, EE+lowP data ($\Lambda$CDM+$r$+$dn_s/d\ln k$) \cite{Planck2015} are specified by gray, red and blue, respectively.}
\label{figure:tachyon}
\end{center}
\end{figure*}

The scalar spectral index and tensor-to-scalar ratio are given by Eqs. (\ref{n_s}) and (\ref{r}), respectively, as follows:
\begin{eqnarray}
\label{n_s,exp}
n_{s} &=& \frac{2N-3}{2N+1},
\\
\label{r,exp}
r &=& \frac{16}{2N+1}.
\end{eqnarray}
{We compute the above equations for the horizon exit $e$-fold number $N_*=50,\, 60$, and present the results in Table \ref{table:exp}. From the table, we infer that larger values of $N_*$ lead to larger $n_s$ and smaller $r$. We also see in Table \ref{table:exp} that the results corresponding to $N_*=50$ are very close to those reported in \cite{Barbosa-Cendejas2015}, where the authors have evaluated the inflationary observables for the exponential potential (\ref{V,exp}) to the second order of slow-roll parameters.}
Furthermore, we can use Eqs. (\ref{n_s,exp}) and (\ref{r,exp}) to plot the $r-n_s$ diagram for the exponential potential (\ref{V,exp}) as shown by a thick black line in Fig. \ref{figure:tachyon}. This plot is drawn for the horizon exit $e$-fold number in the range $50 \leq N_* \leq 60$, and the smaller and larger black points correspond to $N_*=50$ and $N_*=60$, respectively. From Fig. \ref{figure:tachyon}, we conclude that the prediction of the exponential potential (\ref{V,exp}) can lie inside the 95\% CL region of Planck 2015 TT, TE, EE+lowP data \cite{Planck2015}. Our result implies that this potential is consistent with the observational data, if the $e$-fold number of horizon exit is taken in the range $57 \lesssim {N_*} \le 60$. Note that the consistency of the exponential potential (\ref{V,exp}) with the observations in $r-n_s$ plane has been already confirmed in light of the first year WMAP data by \cite{Steer2004}.

\begin{table}
\caption{{Estimated values of inflationary observables for the exponential potential (\ref{V,exp}) with $N_{*}=50,\,60$ in the tachyon inflationary scenario.}}
\label{table:exp}
\scalebox{1}{
\begin{tabular}{|c|c|c|c|c|c|}
\hline
$N_{*}$ & $n_{s}$ & $\frac{dn_{s}}{d\ln k}$ & $n_{t}$ & $r$ & $f_{\mathrm{NL}}^{\mathrm{local}}$\tabularnewline
\hline
50 & 0.9604 & -0.0008 & -0.0198 & 0.1584 & 0.0165\tabularnewline
\hline
60 & 0.9669 & -0.0005 & -0.0165 & 0.1322 & 0.0138\tabularnewline
\hline
\end{tabular}
}
\end{table}

Surprisingly, we see that the prediction of the exponential potential (\ref{V,exp}) for the observables $n_s$ and $r$ in the framework of tachyon inflation coincides with the results of the chaotic quadratic potential $V(\phi)=m^{2}\phi^{2}/2$ in the standard inflationary setting \cite{Planck2015}. In order to explain this surprising result, we note that by changing variables, the tachyon action (\ref{S}) at linear order can be written in the form of standard canonical action \cite{Sen2002c, Steer2004}. In this way, the exponential potential (\ref{V,exp}) in the tachyon framework is transformed to the quadratic potential in the standard canonical setting.

So far, we have shown that the prediction of the exponential potential (\ref{V,exp}) in $r-n_s$ plane can be compatible with the Planck 2015 data \cite{Planck2015}. In the next step, we want to estimate the other inflationary observables including the running of the scalar spectral index $dn_s/d\ln k$, the tensor spectral index $n_t$, and the local non-Gaussianity parameter $f_{{\rm NL}}^{{\rm local}}$, for this potential. Applying Eq. (\ref{V,exp}) in Eqs. (\ref{dn_s/d{ln}k}), (\ref{n_t}), and (\ref{f_{NL}^{local}}), we reach
\begin{eqnarray}
\label{dns,exp}
\frac{dn_{s}}{d\ln k} &=& -\frac{8}{\left(2N+1\right)^{2}},
\\
\label{n_t,exp}
n_{t} &=& -\frac{2}{2N+1},
\\
\label{f_{NL}^{local},exp}
f_{{\rm NL}}^{{\rm local}} &=& \frac{5}{3\left(2N+1\right)}.
\end{eqnarray}
{We have listed the resulting values from the above equations for $N_*=50,\, 60$ in Table \ref{table:exp}.} In addition, in Table \ref{table:tachyon}, we have summarized the lower and upper bounds of $dn_s/d\ln k$ and $f_{{\rm NL}}^{{\rm local}}$ for the parameter space of $N_*$ for which the exponential potential (\ref{V,exp}) is compatible with the Planck 2015 results in $r-n_s$ test. As we see in Tables \ref{table:exp} and \ref{table:tachyon}, the obtained values for $dn_s/d\ln k$ are of order $10^{-4}$, and thus they are consistent with the 95\% CL constraint of Planck 2015 TT, TE, EE+lowP data \cite{Planck2015}. In addition, the evaluated values for $f_{{\rm NL}}^{{\rm local}}$ are of order $10^{-2}$ which are in agreement with the 68\% CL constraint of Planck 2015 T+E data \cite{Planck2015NG}.

\begin{table*}[t]
\caption{Predictions of different steep potentials for the running of the scalar index $dn_s/d\ln k$ and the local non-Gaussianity parameter $f_{\mathrm{NL}}^{\mathrm{local}}$ in the tachyon inflationary scenario.}
\label{table:tachyon}
\scalebox{1}{
\begin{tabular}{ccccccccccc}
\hline
\hline
Potential & $\qquad$ & $\begin{array}{c}
\mathrm{Fixed}\\
\mathrm{parameter}
\end{array}$ & $\qquad$ & $\begin{array}{c}
\mathrm{Varying}\\
\mathrm{parameter}
\end{array}$ & $\qquad$ & $\begin{array}{c}
r-n_{s}\\
\mathrm{consistency}
\end{array}$ & $\qquad$ & $\frac{dn_{s}}{d\ln k}$ & $\qquad$ & $f_{\mathrm{NL}}^{\mathrm{local}}$\tabularnewline
\hline
 &  &  &  &  &  &  &  &  &  & \tabularnewline
$\mathrm{Exponential}$ &  & $-$ &  & $57\lesssim N_{*}\le60$ &  & 95\% CL &  & $[-0.0006,-0.0005]$ &  & $[0.0138,0.0145]$\tabularnewline
 &  &  &  &  &  &  &  &  &  & \tabularnewline
$\begin{array}{c}
\mathrm{Inverse}\\
\mathrm{power-law}
\end{array}$ &  & $N_{*}=60$ &  & $n\gtrsim40$ &  & 95\% CL &  & $[-0.0006,-0.0005]$ &  & $[0.0138,0.0141]$\tabularnewline
 &  &  &  &  &  &  &  &  &  & \tabularnewline
$\mathrm{Inverse}\cosh$ &  & $N_{*}=60$ &  & $x\gtrsim51.5$ &  & 95\% CL &  & $[-0.0005,-0.0003]$ &  & $[0.0138,0.0196]$\tabularnewline
 &  &  &  &  &  &  &  &  &  & \tabularnewline
$\begin{array}{c}
\mathrm{Mutated}\\
\mathrm{exponential}
\end{array}$ &  & $n=4$ &  & $57\lesssim N_{*}\le60$ &  & 95\% CL &  & $[-0.0006,-0.0005]$ &  & $[0.0137,0.0144]$\tabularnewline
 &  &  &  &  &  &  &  &  &  & \tabularnewline
\hline
\end{tabular}
}
\end{table*}

\section{Inverse power-law potential}
\label{section:ipl}

The next steep potential that we investigate in the framework of tachyon inflation, is the inverse power-law potential
\begin{equation}
\label{V,ipl}
V(\phi)=V_{0}\left(\frac{\phi}{\phi_{0}}\right)^{-n},
\end{equation}
where $V_0$ and $n>0$ are constant.
{The asymptotic treatment of this steep potential agrees with the implication of the original string-motivated idea of \cite{Sen2002a, Sen2002b, Sen2002c}, where the rolling tachyon field is employed to describe the low-energy sector for D-branes and open strings \cite{Abramo2003}. Since the shape of this potential goes to zero at infinite values of the field, therefore in this asymptotic vacuum there exists no D-branes, and thus no open strings \cite{Abramo2003}.}
In the framework of standard inflationary scenario, this potential leads to the intermediate inflation with the scale factor $a(t)\propto\exp(At^{f})$ where $A > 0$ and $0 < f = 4/(n + 4) < 1$ \cite{Barrow1990, Barrow1993, Barrow2006, Barrow2007}, which is completely ruled out by the Planck 2015 data \cite{Planck2015}, as it has been shown explicitly in \cite{Rezazadeh2015, Rezazadeh2016}.

In order to examine this potential in the tachyon inflationary scenario, first we calculate the first slow-roll parameter from Eq. (\ref{varepsilon_1}), and obtain
\begin{equation}
\label{varepsilon_1,ipl}
\varepsilon_{1}=\frac{n^{2}}{2V_{0}\phi_{0}^{n}}\phi^{n-2}.
\end{equation}
This equation implies that for the case $n=2$, the first slow-roll parameter $\varepsilon _1$ becomes constant which is related to the power-law inflation with the scale factor $a(t) \propto t^q$ where $q>1$. Therefore, in this case inflation never ends by slow-roll violation, and we encounter a graceful exit problem for the model. Also, for the case $n<2$, $\varepsilon _1$ becomes a decreasing function during inflation and consequently inflation cannot stop by slow-roll violation. But, for the case $n>2$, $\varepsilon _1$ becomes an increasing function, and hence inflation can successfully terminate in this case. In the following subsections, we will investigate these three cases, separately, in ample detail.

\subsection{Case $n=2$}
\label{subsection:n=2,ipl}

For this case, we see easily that Eq. (\ref{varepsilon_1,ipl}) gives the first slow-roll parameter as
\begin{equation}
\label{varepsilon_1,n=2,ipl}
\varepsilon_{1}=\frac{2}{V_{0}\phi_{0}^{2}},
\end{equation}
which is constant. Solving the differential equation (\ref{dot{phi}}), we find the evolution of inflaton versus time in the form
\begin{equation}
\label{phi,t,n=2,ipl}
\phi=\frac{2}{\sqrt{3V_{0}}\,\phi_{0}}\,t,
\end{equation}
where we have set the constant of integration to zero without loss of generality. Using this and the Friedmann equation (\ref{H,V}), the Hubble parameter in the slow-roll regime turns into
\begin{equation}
\label{H,t,n=2,ipl}
H = \frac{q}{t},
\end{equation}
where
\begin{equation}
\label{q}
q\equiv\frac{V_{0}\phi_{0}^{2}}{2}=\frac{1}{\varepsilon_{1}}.
\end{equation}
The Hubble parameter given in Eq. (\ref{H,t,n=2,ipl}) leads to the scale factor
\begin{equation}
 \label{a,t,n=2,ipl}
 a(t)\propto t^{q},
\end{equation}
which is associated with the power-law inflation. Therefore, we conclude that the inverse power-law potential (\ref{V,ipl}) with $n=2$ gives rise to the power-law inflation in the tachyon framework. {We have derived this result in the slow-roll approximation, and it is nevertheless in agreement with the result of the exact analysis of \cite{Padmanabhan2002, Feinstein2002, Abramo2003}.}

Solving the differential equation (\ref{d{phi}/dN}), we find the evolution of inflaton with respect to $N$ as
\begin{equation}
\label{phi,N,n=2,ipl}
\phi=\phi_{e}e^{-\frac{N}{q}},
\end{equation}
where $\phi _e$ indicates the value of inflaton at the end of inflation. The important note is that $\phi _e$ cannot be determined by setting $\varepsilon _1 = 1$, because $\varepsilon _1$ is constant during inflation and never reaches unity. Here, to overcome this problem, we follow the logic of \cite{Martin2014a}, and retain $\phi_e$ as an extra parameter. We can determine it by fixing the amplitude of the scalar power spectrum at the horizon crossing from the observational data. To do so, we use Eq. (\ref{phi,N,n=2,ipl}) in Eq. (\ref{mathcal{P}_s}), and obtain the scalar power spectrum at the time of horizon exit as
\begin{equation}
\label{mathcal{P}_s,n=2,ipl}
{\cal P}_{s}\left(N_{*}\right)=\frac{q^{2}}{12\pi^{2}\phi_{e}^{2}}e^{\frac{2N_{*}}{q}}.
\end{equation}
Solving the above equation for $\phi_e$ yields
\begin{equation}
\label{phi_e,n=2,ipl}
\phi_{e}=\frac{q}{2\pi\sqrt{3{\cal P}_{s}\left(N_{*}\right)}}e^{\frac{N_{*}}{q}}.
\end{equation}

In order to check viability of the model in $r-n_s$ plane, we use Eq. (\ref{phi,N,n=2,ipl}) in Eqs. (\ref{n_s}) and (\ref{r}), and obtain
\begin{eqnarray}
\label{n_s,n=2,ipl}
n_{s} &=& 1-\frac{2}{q},
\\
\label{r,n=2,ipl}
r &=& \frac{16}{q}.
\end{eqnarray}
An important point is that the expressions found for $n_s$ and $r$ in the above equations are same as those obtained in the standard canonical inflation \cite{Unnikrishnan2013, Rezazadeh2016}. We can easily combine these two equations to eliminate the parameter $q$ between them, and reach
\begin{equation}
\label{r,n_s,n=2,ipl}
r=8\left(1-n_{s}\right),
\end{equation}
implying a linear relation between $r$ and $n_s$. Note that this relation is same as that obtained for the power-law inflation in the context of the standard canonical inflation \cite{Tsujikawa2013}, and also the Brans-Dicke inflation \cite{Tahmasebzadeh2016}. It should be noticed that we can combine the equations for $n_s$ and $r$ just to eliminate the constant parameters, and we are not allowed to do this for the dynamical quantities such as time $t$, scalar field $\phi$, or $e$-fold number $N$. Because, otherwise we may mistakenly neglect the important fact that the observables must be evaluated at the time of horizon exit. Now, we can apply Eq. (\ref{r,n_s,n=2,ipl}) to draw $r-n_s$ plot as demonstrated by a solid blue line in Fig. \ref{figure:tachyon}. It is apparent from the figure that the plot of the power-law inflation model in the tachyon framework coincides exactly with the one for this model in the canonical scenario \cite{Planck2015, Unnikrishnan2013, Rezazadeh2016, Rezazadeh2017}, and it lies completely outside the region allowed by the Planck 2015 data \cite{Planck2015}. Consequently, we see that the inverse power-law potential with $n=2$ is ruled out by the current observational data.

\subsection{Case $n<2$}
\label{subsection:n<2,ipl}

As we mentioned above, in this case inflation cannot end by slow-roll violation too. Now we are interested to find the scale factor for this case. For this purpose, we use Eq. (\ref{V,ipl}) in Eq. (\ref{dot{phi}}), and attain
\begin{equation}
\label{phi,t,n<2,ipl}
\phi=\left[\frac{\left[n(4-n)t\right]^{2}}{12V_{0}\phi_{0}^{n}}\right]^{\frac{1}{4-n}},
\end{equation}
where we have considered the integration constant to zero without loss of generality. Now, we apply the above equation in Eq. (\ref{H,V}), and then we will have
\begin{equation}
\label{H,t,n<2,ipl}
H=\left[\frac{2^{n}V_{0}^{2}\phi_{0}^{2n}}{3^{2-n}\left[n(4-n)t\right]^{n}}\right]^{\frac{1}{4-n}}.
\end{equation}
This Hubble parameter leads to the intermediate scale factor
\begin{equation}
\label{a,t,n<2,ipl}
a(t)\propto\exp\left(At^{f}\right),
\end{equation}
where the parameters $A$ and $f$ are defined as
\begin{eqnarray}
\label{A}
A &=& \frac{1}{2-n}\left[\frac{(4-n)^{4-2n}V_{0}^{2}\phi_{0}^{2n}}{12^{2-n}n^{n}}\right]^{\frac{1}{4-n}},
\\
\label{f}
f &=& \frac{4-2n}{4-n}.
\end{eqnarray}
It is evident that $A>0$ and $0<f<1$, and hence the required conditions for having an intermediate scale factor are satisfied. Therefore, we see that within the framework of tachyon inflation, the inverse power-law potential (\ref{V,ipl}) with $n<2$ gives rise to the intermediate inflation. {This verifies the results obtained in \cite{Feinstein2002, Abramo2003}.}

Here, we turn to estimate the observational quantities for this model. First we solve the differential equation (\ref{d{phi}/dN}) for the potential (\ref{V,ipl}) and obtain
\begin{equation}
\label{phi,N,n<2,ipl}
\phi=\left[\phi_{e}^{2-n}-\frac{n(2-n)}{V_{0}\phi_{0}^{n}}N\right]^{\frac{1}{2-n}},
\end{equation}
where $\phi_e$ denotes the inflaton  at the end of inflation. Since the first slow-roll parameter $\varepsilon _1$ is a decreasing function and cannot reach unity at the end of inflation, then we cannot determine $\phi_e$ by setting $\varepsilon _1=1$. Thus, we again follow the procedure of \cite{Martin2014a} to keep it as an extra parameter in the above equation. To determine it, we repeat the procedure applied in the previous subsection and fix the amplitude of the scalar perturbations from the observational data. Using Eqs. (\ref{mathcal{P}_s}) and (\ref{phi,N,n<2,ipl}), we have
\begin{equation}
\label{mathcal{P}_s,n<2,ipl}
{\cal P}_{s}\left(N_{*}\right)=\frac{1}{12}\left(\frac{V_{0}\phi_{0}^{n}}{\pi n}\right)^{2}\left[\phi_{e}^{2-n}-\frac{n(2-n)}{V_{0}\phi_{0}^{n}}N_{*}\right]^{\frac{2-2n}{2-n}}.
\end{equation}
We  recall  that the scalar power spectrum at the horizon exit can be fixed as
$\ln\left[10^{10}{\cal P}_{s}\left(N_{*}\right)\right]=3.094\pm0.034$ from Planck 2015 TT, TE, EE+lowP data \cite{Planck2015}. The above equation can be simply solved for $\phi_e$ to give
\begin{equation}
\label{phi_e,n<2,ipl}
\phi_{e}=\left[\left(\frac{2\pi n\sqrt{3{\cal P}_{s}\left(N_{*}\right)}}{V_{0}\phi_{0}^{n}}\right)^{\frac{2-n}{1-n}}+\frac{n(2-n)}{V_{0}\phi_{0}^{n}}N_{*}\right]^{\frac{1}{2-n}}.
\end{equation}
Now, by the
 use of Eqs. (\ref{phi,N,n<2,ipl}) and (\ref{phi_e,n<2,ipl}), we see that Eqs. (\ref{n_s}) and (\ref{r}), respectively, lead to
\begin{eqnarray}
\label{n_s,n<2,ipl}
n_{s} &=& 1-\left(n-1\right)\left[\frac{2n\left(\pi\sqrt{3{\cal P}_{s}\left(N_{*}\right)}\right)^{2-n}}{V_{0}\phi_{0}^{n}}\right]^{\frac{1}{n-1}},
\\
\label{r,n<2,ipl}
r &=& \left[\frac{2^{2n-1}n^{n}\left(\pi\sqrt{3{\cal P}_{s}\left(N_{*}\right)}\right)^{2-n}}{V_{0}\phi_{0}^{n}}\right]^{\frac{1}{n-1}}.
\end{eqnarray}
It is evident from Eq. (\ref{n_s,n<2,ipl}) that for $n=1$, we have $n_s=1$ referred as the scale-invariant Harrison-Zel'dovich spectrum which is completely ruled out by the Planck 2015 data \cite{Planck2015}. For $n<1$, we have a red-tilted scalar spectrum ($n_s<1$), while $n>1$ leads to a blue-tilted spectrum ($n_s>1$) being inconsistent with the Planck 2015 data \cite{Planck2015}. We see surprisingly that $n_s$ and $r$ in the above equations do not depend on the dynamical variables such as $t$, $\phi$, or $N$, at all. Therefore, we can combine these two equations to get
\begin{equation}
\label{r,n_s,n<2,ipl}
r=\frac{4n}{n-1}\left(1-n_{s}\right),
\end{equation}
indicating a linear relation between $r$ and $n_s$. It is obvious that for $n \to 2$, this equation reduces to the relation (\ref{r,n_s,n=2,ipl}) obtained for the power-law inflation in the previous subsection. Using Eq. (\ref{r,n_s,n<2,ipl}), the $r-n_s$ diagram for the model can be plotted as shown by a dashed blue lines in Fig. \ref{figure:tachyon}, that each line is related to the specified value of $n$. We see in the figure that as the parameter $n$ approaches $2$, the prediction of intermediate inflation approaches the result of power-law inflation. This is just the behavior expected from Eq. (\ref{r,n_s,n<2,ipl}), as we mentioned above. From Fig. \ref{figure:tachyon}, we conclude that although the intermediate inflation driven by the inverse power-law potential (\ref{V,ipl}) with $n<2$ can be compatible with the Planck 2013 results (see e.g. $n=1.9$), it is completely ruled out by the Planck 2015 data. Note that the consistency of the intermediate inflation within the tachyon scenario in light of the Planck 2013 data has been already confirmed by \cite{Nozari2013}.

\subsection{Case $n>2$}
\label{subsection:n>2,ipl}

In this subsection, we concentrate on examination of the inverse power-law potential (\ref{V,ipl}) with $n>2$. {For $n\neq4$, the temporal evolution of the inflaton field is still given by Eq. (\ref{phi,t,n<2,ipl}). As a result, the scale factor is given by Eq. (\ref{a,t,n<2,ipl}) yet, but now $A<0$, as it is clear from Eq. (\ref{A}). Furthermore, from Eq. (\ref{f}), we see that $f<0$ and $f>2$ for $n<4$ and $n>4$, respectively. This means that the scale factor (\ref{a,t,n<2,ipl}) is no longer intermediate.}

{For $n=4$, the temporal evolution of the inflaton results from Eq. (\ref{dot{phi}}) as
\begin{equation}
 \label{phi,t,n>2,ipl}
\phi=\phi_{i}e^{\frac{4t}{\sqrt{3V_{0}}\,\phi_{0}^{2}}},
\end{equation}
where $\phi_{i}\equiv\phi(t=0)$ is the constant of integration. By use of this result in the slow-roll Friedmann equation (\ref{H,V}), we find
\begin{equation}
 \label{H,t,n=4,ipl}
 H=\sqrt{\frac{V_{0}}{3}}\left(\frac{\phi_{0}}{\phi_{i}}\right)^{2}e^{-\frac{8t}{\sqrt{3V_{0}}\,\phi_{0}^{2}}}.
\end{equation}
Consequently, the above equation gives the scale factor as
\begin{equation}
 \label{a,t,n=4,ipl}
 a\propto\exp\left[\frac{V_{0}\phi_{0}^{4}}{8\phi_{i}^{2}}\left(1-e^{-\frac{8t}{\sqrt{3V_{0}}\,\phi_{0}^{2}}}\right)\right].
\end{equation}
Additionally, using Eq. (\ref{H,t,n=4,ipl}) in (\ref{varepsilon_1}), the first slow-roll parameter (\ref{varepsilon_1}) versus time becomes
\begin{equation}
 \label{varepsilon_1,t,n=4,ipl}
 \varepsilon_{1}=\frac{8\phi_{i}^{2}}{V_{0}\phi_{0}^{4}}e^{\frac{8t}{\sqrt{3V_{0}}\,\phi_{0}^{2}}},
\end{equation}
which explicitly shows that $\varepsilon_{1}$ increases with time during inflation in the case $n=4$.}

{Although, the evolutionary behavior of the inflaton $\phi$ versus time $t$ is different in the cases $n=4$ and $n\neq4$, but its evolution versus $e$-fold number $N$ is same in the both cases, and it is still given by Eq. (\ref{phi,N,n<2,ipl}). However, now it is possible to determine $\phi_e$ by setting $\varepsilon _1 = 1$ in Eq. (\ref{varepsilon_1,ipl}), because as we see from this equation, in the case $n>2$, the first slow-roll parameter is an increasing function during inflation. In this way, we find
\begin{equation}
\label{phi_e,n>2,ipl}
\phi_{e}=\left(\frac{2V_{0}\phi_{0}^{n}}{n^{2}}\right)^{\frac{1}{n-2}}.
\end{equation}
}
If we use the above results together with Eq. (\ref{phi,N,n<2,ipl}) in Eq. (\ref{mathcal{P}_s}), then we will have
\begin{equation}
\label{mathcal{P}_s,n>2,ipl}
{\cal P}_{s}=\frac{1}{3\pi^{2}}\left[\frac{n}{2^{2n-3}V_{0}\phi_{0}^{n}}\big(2(n-2)N+n\big)^{n-1}\right]^{\frac{2}{n-2}}.
\end{equation}
Fixing ${\cal P}_s$ at the horizon exit from the observational data, the above relation can be applied to determine $V_0$ or $\phi_0$ in terms of the other parameters of the model.

Moreover, by the
 use of Eqs. (\ref{phi,N,n<2,ipl}) and (\ref{phi_e,n>2,ipl}) in Eqs. (\ref{n_s}) and (\ref{r}), we obtain
\begin{eqnarray}
\label{n_s,n>2,ipl}
n_{s} &=& 1-\frac{4\left(n-1\right)}{2\left(n-2\right)N+n},
\\
\label{r,n>2,ipl}
r &=& \frac{16n}{2\left(n-2\right)N+n}.
\end{eqnarray}
{Here, we can use the above equations to estimate the observables $n_s$ and $r$ for the inverse power-law potential (\ref{V,ipl}). The obtained results for some typical values of the parameter $n$ has been listed in Table \ref{table:ipl}. From the table, it is apparent that as $n$ increases, then $n_s$ increases and $r$ decreases. Also, comparing Table \ref{table:ipl} with Table \ref{table:exp}, we infer that for $n\gg1$, the prediction of the inverse power-law potential (\ref{V,ipl}) approaches the results of the exponential potential (\ref{V,exp}).}
The obtained relations in two Eqs. (\ref{n_s,n>2,ipl}) and (\ref{r,n>2,ipl}) also can be utilized to make the $r-n_s$ plot for the model. This plot is demonstrated in Fig. \ref{figure:tachyon} for $N_*=50$ and $N_*=60$ by the dashed and solid red lines, respectively. To make these plots we have considered $n$ as the varying parameter. From the figure, we conclude that with $N_*=50$, the inverse power-law potential (\ref{V,ipl}) cannot be compatible with the Planck 2015 observational results \cite{Planck2015}. But, our computations show that with $N_*=60$, if the parameter $n$ varies in the range $n \gtrsim 40$, then the result of this potential (\ref{V,ipl}) can place within the 95\% CL region of Planck 2015 TT, TE, EE+lowP data \cite{Planck2015}.

\begin{table}
\caption{{Estimated values of the inflationary observables for the inverse power-law potential (\ref{V,ipl}) with $n>2$, for some typical values of the parameter $n$ in the tachyon inflationary framework.}}
\label{table:ipl}
\scalebox{1}{
\begin{tabular}{|c|c|c|c|c|c|c|}
\hline
$n$ & $N_{*}$ & $n_{s}$ & $\frac{dn_{s}}{d\ln k}$ & $n_{t}$ & $r$ & $f_{\mathrm{NL}}^{\mathrm{local}}$\tabularnewline
\hline
\multirow{2}{*}{4} & 50 & 0.9412 & -0.0012 & -0.0392 & 0.3137 & 0.0245\tabularnewline
\cline{2-7}
 & 60 & \multirow{1}{*}{0.9508} & -0.0008 & -0.0328 & 0.2623 & 0.0205\tabularnewline
\hline
\multirow{2}{*}{40} & 50 & 0.9594 & -0.0008 & -0.0208 & 0.1667 & 0.0169\tabularnewline
\cline{2-7}
 & 60 & 0.9661 & -0.0006 & -0.0174 & 0.1391 & 0.0141\tabularnewline
\hline
\multirow{2}{*}{400} & 50 & 0.9603 & -0.0008 & -0.0199 & 0.1592 & 0.0165\tabularnewline
\cline{2-7}
 & 60 & 0.9669 & -0.0005 & -0.0166 & 0.1329 & 0.0138\tabularnewline
\hline
\end{tabular}
}
\end{table}

Now, we proceed to evaluate the other inflationary observables for this model. Using Eqs. (\ref{phi,N,n<2,ipl}) and (\ref{phi_e,n>2,ipl}), it can be shown that Eqs. (\ref{dn_s/d{ln}k}), (\ref{n_t}), and (\ref{f_{NL}^{local}}) give the running of the scalar spectral index, the tensor spectral index, and the local non-Gaussianity parameter, respectively, as follows:
\begin{eqnarray}
\label{dn_s/d{ln}k,n>2,ipl}
\frac{dn_{s}}{d\ln k} &=& -\frac{8\left(n-1\right)\left(n-2\right)}{\left[2\left(n-2\right)N+n\right]^{2}},
\\
\label{n_t,n>2,ipl}
n_{t} &=& -\frac{2n}{2\left(n-2\right)N+n},
\\
\label{f_{NL}^{local},n>2,ipl}
f_{{\rm NL}}^{{\rm local}} &=& \frac{5\left(n-1\right)}{3\left[2\left(n-2\right)N+n\right]}.
\end{eqnarray}
{These observables have been estimated in Table \ref{table:ipl} for some typical values of $n$.}
Also, we have specified the intervals of observables $d{n_s}/d\ln k$ and $f_{{\rm NL}}^{{\rm local}}$ for the range $n \gtrsim 40$ in Table \ref{table:tachyon}. The evaluated values for $d{n_s}/d\ln k$ are in agreement with the 95\% CL range of Planck 2015 TT, TE, EE+lowP data \cite{Planck2015}. Also the obtained values for $f_{{\rm NL}}^{{\rm local}}$ agree with the 68\% CL constraints of Planck 2015 T+E data \cite{Planck2015NG}.

{Here, we emphasize that in the cases $n=2$ (power-law inflation) and $n<2$ (intermediate inflation) of the inverse power-law potential (\ref{V,ipl}), the slow-roll conditions are not violated, and hence the Universe cannot transit to the decelerating phase of the radiation-dominated era. Also, in the case $n>2$ (dust), although the slow-roll conditions can be violated around the end time of inflation, but the tachyon energy density remains as a dominated component in the energy density of the Universe, which is not favored. To overcome these problems, it is possible to add a reheating process to the final stages of the inflationary period, that makes it possible for the Universe to enter the radiation-dominated era, but its details are unknown so far. We recall that the mechanism of the conventional reheating process is not applicable for the tachyon scenario \cite{Kofman2002}, and one should consider alternative mechanisms.}

\section{Inverse cosh potential}
\label{section:icosh}

In this section, we will focus on the inverse cosh potential \cite{Leblond2003, Lambert2007, Kim2003, Steer2004, Barbosa-Cendejas2015}
\begin{equation}
\label{V,icosh}
V(\phi)=\frac{V_{0}}{\cosh\left(\frac{\phi}{\phi_{0}}\right)},
\end{equation}
where $V_0$ and $\phi_0$ are constant parameters of the model. This potential has extensive applications in study of unstable D-branes in the context of string theory \cite{Leblond2003, Lambert2007}. Application of this potential in cosmological contexts appears in \cite{ Steer2004, Barbosa-Cendejas2015, Kim2003}.

We start with calculating the first slow-roll parameter in Eq. (\ref{varepsilon_1,V}), and find
\begin{equation}
\label{varepsilon_1,icosh}
\varepsilon_{1}=\frac{1}{2V_{0}\phi_{0}^{2}}\sinh\left(\frac{\phi}{\phi_{0}}\right)\tanh\left(\frac{\phi}{\phi_{0}}\right).
\end{equation}
The above equation clears that $\varepsilon _1$ increases during inflation. Therefore, it is possible to set $\varepsilon _1=1$ to find the inflaton at the end of inflation as
\begin{equation}
\label{phi_e,icosh}
\phi_{e}=\phi_{0}\ln\left[\sqrt{2x\left(\sqrt{x^{2}+1}+x\right)}+\sqrt{x^{2}+1}+x\right],
\end{equation}
where we have defined
\begin{equation}
\label{x}
x\equiv V_{0}\phi_{0}^{2}.
\end{equation}
Solving the differential equation (\ref{d{phi}/dN}) for the potential (\ref{V,icosh}), we find
\begin{eqnarray}
\phi= && 2\phi_{0}\,{\rm arccoth}\Big\{ \coth\Big[\frac{1}{2}\ln\Big(x+\sqrt{x^{2}+1}
\nonumber\\
&&  +\sqrt{2x\left(\sqrt{x^{2}+1}+x\right)}  \Big)\Big]e^{\frac{N}{x}}\Big\},
 \label{phi,N,icosh}
\end{eqnarray}
where we have used $\phi\left(N=0\right)=\phi_{e}$ as the initial condition.

We see that using (\ref{mathcal{P}_s}), the scalar power spectrum for this potential reads
\begin{equation}
\label{mathcal{P}_s,icosh}
{\cal P}_{s}=\frac{x^{2}}{12\pi^{2}\phi_{0}^{2}}\frac{1}{\sinh^{2}\left(\frac{\phi}{\phi_{0}}\right)}.
\end{equation}
One can fix this equation according to the Planck 2015 observational results \cite{Planck2015} to determine one of the parameters $V_0$ or $\phi_0$ in terms of the other parameters of the model.

Also, Eqs. (\ref{n_s}) and (\ref{r}), give the scalar spectral index and the tensor-to-scalar ratio for the inverse $\cosh$ potential (\ref{V,icosh}) in the following forms:
\begin{eqnarray}
\label{n_s,icosh}
n_{s} &=& 1-\frac{2}{x}\cosh\left(\frac{\phi}{\phi_{0}}\right),
\\
\label{r,icosh}
r &=& \frac{8}{x}\sinh\left(\frac{\phi}{\phi_{0}}\right)\tanh\left(\frac{\phi}{\phi_{0}}\right).
\end{eqnarray}
{The above equations have been used to estimate $n_s$ and $r$ for several values of $x$, and the results have been presented in Table \ref{table:icosh}. As we see in the table, by increasing $x$, the values of both $n_s$ and $r$ increase. Also, for $x\gg1$, the prediction of the inverse $\cosh$ potential (\ref{V,icosh}) for these observables approaches the ones of the exponential potential (\ref{V,exp}).}
The above equations also be used to plot the prediction of the model in $r-n_s$ plane as shown by dashed and solid green lines in Fig. \ref{figure:tachyon}. In should be noted that the dashed and solid lines correspond to $N_*=50$ and $N_*=60$, respectively. To plot this diagram, we have considered $x$ as the varying parameter. Our computations indicate that for $N_*=60$, the prediction of the model lies inside the 95\% CL region of Planck 2015 TT, TE, EE+lowP data \cite{Planck2015}, if the parameter $x$ varies in the range $x \gtrsim 51.5$.
{Our result for the inverse $\cosh$ potential (\ref{V,icosh}) in $r-n_s$ plane is in agreement with the one presented in \cite{Barbosa-Cendejas2015}.}
Also, it is worthwhile to mention that the compatibility of the potential (\ref{V,icosh}) with the observational results in $r-n_s$ plane has been already verified in \cite{Steer2004} by use of the first year WMAP data.

\begin{table}
\caption{{Estimated values of inflationary observables for the inverse $\cosh$ potential (\ref{V,icosh}) with some typical values of $x\equiv V_{0}\phi_{0}^{2}$, in the tachyon inflationary scenario.}}
\label{table:icosh}
\scalebox{1}{
\begin{tabular}{|c|c|c|c|c|c|c|}
\hline
$x$ & $N_{*}$ & $n_{s}$ & $\frac{dn_{s}}{d\ln k}$ & $n_{t}$ & $r$ & $f_{\mathrm{NL}}^{\mathrm{local}}$\tabularnewline
\hline
\multirow{2}{*}{$40$} & 50 & 0.9413 & -0.0005 & -0.0081 & 0.0645 & 0.0245\tabularnewline
\cline{2-7}
 & 60 & \multirow{1}{*}{0.9449} & -0.0003 & -0.0049 & 0.0389 & 0.0230\tabularnewline
\hline
\multirow{2}{*}{$10^{2}$} & 50 & 0.9571 & -0.0007 & -0.0168 & 0.1344 & 0.0179\tabularnewline
\cline{2-7}
 & 60 & 0.9630 & -0.0005 & -0.0131 & 0.1047 & 0.0154\tabularnewline
\hline
\multirow{2}{*}{$10^{3}$} & 50 & 0.9604 & -0.0008 & -0.0198 & 0.1581 & 0.0165\tabularnewline
\cline{2-7}
 & 60 & 0.9669 & -0.0005 & -0.0165 & 0.1319 & 0.0138\tabularnewline
\hline
\end{tabular}
}
\end{table}

Here, we use Eq. (\ref{phi,N,icosh}) in Eqs. (\ref{dn_s/d{ln}k}), (\ref{n_t}), and (\ref{f_{NL}^{local}}), and obtain
\begin{eqnarray}
\label{dn_s/d{ln}k,icosh}
\frac{dn_{s}}{d\ln k} &=& -\frac{2}{x}\sinh^{2}\left(\frac{\phi}{\phi_{0}}\right),
\\
\label{n_t,icosh}
n_{t} &=& -\frac{1}{x}\sinh\left(\frac{\phi}{\phi_{0}}\right)\tanh\left(\frac{\phi}{\phi_{0}}\right),
\\
\label{f_{NL}^{local},icosh}
f_{{\rm NL}}^{{\rm local}} &=& \frac{5}{6x}\cosh\left(\frac{\phi}{\phi_{0}}\right).
\end{eqnarray}
{The obtained results from these equations are available in Table \ref{table:icosh} for some typical values of $x$.}
{Moreover, the above equation have been used to estimate $dn_s/d\ln k$ and $f_{{\rm NL}}^{{\rm local}}$ in Table \ref{table:tachyon} for the range of $x$ that the model in hand is successful in $r-n_s$ test.}
The estimated values for these observables are compatible with those deduced from the Planck 2015 data \cite{Planck2015}.

\section{Mutated exponential potential}
\label{section:mexp}

Another steep potential that has been introduced in \cite{Steer2004} is given by
\begin{equation}
\label{V,mexp}
V(\phi)=V_{0}\left[1+\left(n-1\right)^{-\left(n-1\right)}\left(\frac{\phi}{\phi_{0}}\right)^{n}\right]e^{-\frac{\phi}{\phi_{0}}},
\end{equation}
where $V_0$ and $n\geq 1$ are constant. This potential can be considered as a mutated form of the exponential potential (\ref{V,exp}), because it contains an exponential term multiplied by another term. This steep potential satisfies the conditions (\ref{V,tachyon}) required for tachyon potentials.
{From theoretical point of view, it should be noted that this potential can be regarded as a generalized form of the potential $V(\phi)=V_{0}\left(1+\frac{\phi}{\phi_{0}}\right)e^{-\frac{\phi}{\phi_{0}}}$ related to the case $n=1$, and has well-based motivations from background-independent open string theory \cite{Witten1992, Witten1993, Li1993}. The function $V(\phi)=V_{0}\left(1+\frac{\phi}{\phi_{0}}\right)e^{-\frac{\phi}{\phi_{0}}}$ leads to exact tree level tachyon potential when all other fields (corrections associated with derivatives of tachyon and all other fields) vanish in a specific coordinate system whereas regularization is imposed by world-sheet boundary sigma model approach. The perturbative vacuum of this function corresponds to $\phi = 0$,  while the stable
vacuum to which the tachyon condenses is at $\phi = \infty$, where $V(\phi)\to 0$.}
It should be noticed that inflation with the case $n=4$ of the potential (\ref{V,mexp}) has been studied before in \cite{Steer2004}, where the authors have provided the $r-n_s$ diagram of the model and concluded that it can lead to very small values for $r$. Therefore, it is tempting to examine the case $n=4$ of this potential, but with a different approach in which we fix the amplitude of the scalar power spectrum ${\cal P}_s$ from the observational data.

The first slow-roll parameter (\ref{varepsilon_1,V}) for the mutated exponential potential (\ref{V,mexp}) with $n=4$, turns into
\begin{equation}
\label{varepsilon_1,mexp}
\varepsilon_{1}=\frac{27\phi_{0}^{2}\left(\phi-3\phi_{0}\right)^{4}\left(\phi^{2}+2\phi_{0}\phi+3\phi_{0}^{2}\right)^{2}e^{\frac{\phi}{\phi_{0}}}}{2V_{0}\left(\phi^{4}+27\phi_{0}^{4}\right)^{3}}.
\end{equation}
Also, the differential equation (\ref{d{phi}/dN}) takes the form
\begin{equation}
\label{d{phi}/dN,mexp}
\frac{d\phi}{dN}=-\frac{27\phi_{0}^{3}\left(\phi-3\phi_{0}\right)^{2}\left(\phi^{2}+2\phi_{0}\phi+3\phi_{0}^{2}\right)e^{\frac{\phi}{\phi_{0}}}}{V_{0}\left(\phi^{4}+27\phi_{0}^{4}\right)^{2}}.
\end{equation}
Using Eq. (\ref{mathcal{P}_s}), the scalar power spectrum reads
\begin{equation}
\label{mathcal{P}_s,mexp}
{\cal P}_{s}=\frac{V_{0}^{2}\left(\phi^{4}+27\phi_{0}^{4}\right)^{4}e^{-\frac{2\phi}{\phi_{0}}}}{8748\pi^{2}\phi_{0}^{6}\left(\phi-3\phi_{0}\right)^{4}\left(\phi^{2}+2\phi_{0}\phi+3\phi_{0}^{2}\right)^{2}}.
\end{equation}
It is apparent that it is difficult to solve the above three equations analytically, and hence we have to invoke a numerical method for this purpose. In our numerical approach, we first solve $\varepsilon_1=1$ in Eq. (\ref{varepsilon_1,mexp}) and find $\phi_e$. Then, we use $\phi_e$ as the initial condition for solving the differential equation (\ref{d{phi}/dN,mexp}) and obtain $\phi_*$ at the horizon exit with the given $e$-fold number $N_*$. Subsequently, using $\phi_*$ in Eq. (\ref{mathcal{P}_s,mexp}) and fixing the amplitude of the scalar perturbations from the Planck 2015 observational results \cite{Planck2015}, one can determine the parameter $V_0$ in terms of $\phi_0$.

Applying Eqs. (\ref{n_s}) and (\ref{r}), we have
\begin{eqnarray}
n_{s}=&& 1-\frac{54\phi_{0}^{2}e^{\frac{\phi}{\phi_{0}}}}{V_{0}\left(\phi^{4}+27\phi_{0}^{4}\right)^{3}}\Big(\phi^{8}-8\phi_{0}\phi^{7}+20\phi_{0}^{2}\phi^{6}
\nonumber
\\
&& +54\phi_{0}^{4}\phi^{4} -216\phi_{0}^{5}\phi^{3} -324\phi_{0}^{6}\phi^{2}+729\phi_{0}^{8}\Big),
\label{n_s,mexp}
 \\
r= && \frac{216\phi_{0}^{2}\left(\phi -3\phi_{0}\right)^{4}\left(\phi^{2} +2\phi_{0}\phi+3\phi_{0}^{2}\right)^{2}e^{\frac{\phi}{\phi_{0}}}}{V_{0}\left(\phi^{4}+27\phi_{0}^{4}\right)^{3}}.
\label{r,mexp}
\end{eqnarray}
Note that although the parameter $\phi_0$ appears explicitly in the above relations, when we evaluate $\phi_*$ and $V_0$ in the numerical solution, the values of the parameters are combined such that the final results of $n_s$ and $r$ have no dependence to the parameter $\phi_0$, at all. Therefore, $n_s$ and $r$ depend only on the parameter $N_*$.
{The results of Eqs. (\ref{n_s,mexp}) and (\ref{r,mexp}) for $N_{*}=50,\,60$ are listed in Table \ref{table:mexp}. These results show that the predictions of the mutated exponential (\ref{V,mexp}) for $n_s$ and $r$ is very close to those obtained for the exponential potential (\ref{V,exp}).}
The $r-n_s$ diagram of the potential (\ref{V,mexp}) for the range $50\le N_{*}\le60$ has been plotted by a dashed orange line in Fig. \ref{figure:tachyon}. We see in the figure that the tachyon inflationary model with the mutated exponential (\ref{V,mexp}) can be consistent with Planck 2015 TT, TE, EE+lowP data \cite{Planck2015} at 95\% CL. Our computations implies that the model is compatible with the observational data for the range $57 \lesssim N_* \le 60$. Note that although the authors of \cite{Steer2004} have already shown that the results of the mutated exponential potential (\ref{V,mexp}) for $n=4$ are compatible with the first year WMAP observations, they did not fix ${\cal P}_s$ in their calculations. But in our numerical approach, we have fixed ${\cal P}_s$ at the epoch of horizon crossing according to the Planck 2015 observational results \cite{Planck2015}, and in this way we determined the parameter $V_0$. Consequently, we found a much limited range for the prediction of the model in $r-n_s$ plane relative to \cite{Steer2004}.

For the mutated exponential potential (\ref{V,mexp}) with $n=4$, the running of the scalar spectral index (\ref{dn_s/d{ln}k}), the tensor spectral index (\ref{n_t}), and the local non-Gaussianity parameter (\ref{f_{NL}^{local}}) turn into
\begin{eqnarray}
\frac{dn_{s}}{d\ln k}=&& -\frac{1458\phi_{0}^{4}\left(\phi-3\phi_{0}\right)^{2}\left(\phi^{2}+2\phi_{0}\phi+3\phi_{0}^{2}\right)e^{\frac{2\phi}{\phi_{0}}}}{V_{0}^{2}\left(\phi^{4}+27\phi_{0}^{4}\right)^{6}}
\nonumber
\\
&& \times \Big(\phi^{12}-12\phi_{0}\phi^{11}+60\phi_{0}^{2}\phi^{10}-120\phi_{0}^{3}\phi^{9}
\nonumber
\\
&& +81\phi_{0}^{4}\phi^{8}-648\phi_{0}^{5}\phi^{7}+648\phi_{0}^{6}\phi^{6}+6480\phi_{0}^{7}\phi^{5}
\nonumber
\\
&& +2187\phi_{0}^{8}\phi^{4}-8748\phi_{0}^{9}\phi^{3}-26244\phi_{0}^{10}\phi^{2}
\nonumber
\\
&& -17496\phi_{0}^{11}\phi+19683\phi_{0}^{12}\Big),
\label{dn_s/d{ln}k,mexp}
\\
n_{t} = && -\frac{27\phi_{0}^{2}\left(\phi^{2}+2\phi_{0}\phi+3\phi_{0}^{2}\right)^{2}\left(\phi-3\phi_{0}\right)^{4}e^{\frac{\phi}{\phi_{0}}}}{V_{0}\left(\phi{}^{4}+27\phi_{0}^{4}\right)^{3}},
\nonumber
\\
\label{n_t,mexp}
\\
f_{{\rm NL}}^{{\rm local}}= && \frac{45\phi_{0}^{2}e^{\frac{\phi}{\phi_{0}}}}{2V_{0}\left(\phi^{4}+27\phi_{0}^{4}\right)^{3}}\Big(\phi^{8}-8\phi_{0}\phi^{7}+20\phi_{0}^{2}\phi^{6}
\nonumber
\\
&& +54\phi_{0}^{4}\phi^{4} -216\phi_{0}^{5}\phi^{3}-324\phi_{0}^{6}\phi^{2}+729\phi_{0}^{8}\Big).
\nonumber
\\
\label{f_{NL}^{local},mexp}
\end{eqnarray}
{These observables have been estimated in Table \ref{table:mexp} for the horizon exit $e$-fold numbers $N_{*}=50,\,60$.}
In Table \ref{table:tachyon}, we also have specified the range of observables $d{n_s}/d\ln k$ and $f_{{\rm NL}}^{{\rm local}}$ in the interval $57 \lesssim N_* \le 60$ for which the mutated exponential potential (\ref{V,mexp}) with $n=4$ is consistent with the 95\% CL constraints of the Planck 2015 data \cite{Planck2015} in $r-n_s$ plane. The obtained values for these observables are compatible with the Planck 2015 results \cite{Planck2015}.

\begin{table}
\caption{{Estimated values of inflationary for the mutated exponential potential (\ref{V,mexp}) with $n=4$, and considering $N_{*}=50,\,60$ in the tachyon inflationary scenario.}}
\label{table:mexp}
\scalebox{1}{
\begin{tabular}{|c|c|c|c|c|c|}
\hline
$N_{*}$ & $n_{s}$ & $\frac{dn_{s}}{d\ln k}$ & $n_{t}$ & $r$ & $f_{\mathrm{NL}}^{\mathrm{local}}$\tabularnewline
\hline
50 & 0.9606 & -0.0008 & -0.0196 & 0.1566 & 0.0164\tabularnewline
\hline
60 & 0.9671 & -0.0005 & -0.0163 & 0.1306 & 0.0137\tabularnewline
\hline
\end{tabular}
}
\end{table}

\section{Conclusions}
\label{section:conclusions}

We studied tachyon inflation with several steep potentials in light of the Planck 2015 data \cite{Planck2015}. For this purpose, we first obtained the necessary relations governing the inflationary observables containing the scalar spectral index $n_s$, the tensor-to-scalar ratio $r$, the running of the scalar spectral index $dn_s/d \ln k$, and the local non-Gaussianity parameter $f_{{\rm NL}}^{{\rm local}}$ in the slow-roll approximation.

In the next step, we checked the consistency of various steep potentials containing exponential, inverse power-law, inverse cosh and mutated exponential potentials with the current observational data of Planck 2015. We depicted the predictions of the potentials in $r-n_s$ plane and obtained the allowed range of the model parameters.

For the exponential potential $V(\phi)=V_{0}e^{-\phi/\phi_{0}}$, we concluded that it is compatible with the 95\% CL region of Planck 2015 TT, TE, EE+lowP data \cite{Planck2015}, if we take the range of the $e$-fold number of the horizon exit as $57 \lesssim N_* \le 60$.

For the inverse power-law potential $V(\phi)=V_{0}(\phi/\phi_{0})^{-n}$, we examined the three cases $n=2$, $n<2$, and $n>2$, separately. In the case $n=2$, this potential leads to the power-law inflation with the scale factor $a(t) \propto t^q$ where $q>1$. An important result is that the $r-n_s$ plot of the power-law inflation model in the tachyon framework coincides with the one for this model in the canonical scenario, and thus it places completely outside the region favored by the Planck 2015 data \cite{Planck2015}. In the case $n<2$, we showed clearly that the inverse power-law potential gives rise to the intermediate inflation with the scale factor $a(t)\propto\exp\left(At^{f}\right)$ where $A>0$ and $0<f<1$. We found that the prediction of this model in $r-n_s$ plane lies entirely outside the region allowed by the Planck 2015 data \cite{Planck2015}. In the case $n>2$, in contrast with the two cases $n=2$ and $n<2$, inflation can end by slow-roll violation. In this case, the potential can be compatible with Planck 2015 TT, TE, EE+lowP data \cite{Planck2015} at 95\% CL, if we take the $e$-fold number of horizon exit as $N_*=60$ and the parameter $n$ is chosen in the range $n \gtrsim 40$.

Then, we turned to examine viability of the tachyon inflation with the inverse $\cosh$ potential $V(\phi ) = V_0/\cosh (\phi /\phi _0)$, which has remarkable importance in study of unstable D-branes in the context of string theory. We concluded that the potential prediction in $r-n_s$ plane lies inside the 95\% CL of Planck 2015 TT, TE, EE+lowP data \cite{Planck2015}, if we consider $N_*=60$ and $x \equiv V_0 \phi _0^2 \gtrsim 51.5$.

The last steep potential that we investigated in the tachyon inflationary scenario, was the mutated exponential potential $V(\phi)=V_{0}[1+\left(n-1\right)^{-(n-1)}(\phi/\phi_{0})^{n}]e^{-\phi/\phi_{0}}$. We focused on the case $n=4$ of this potential and found that for $57 \lesssim {N_*} \le 60$, the $r-n_s$ plot of the model is consistent with the 95\% CL region of Planck 2015 TT, TE, EE+lowP data \cite{Planck2015}.

Besides the $r-n_s$ test examined for the above steep potentials, we also estimated the other inflationary observables including the running of the scalar spectral index $d{n_s}/d\ln k$ and the local non-Gaussianity parameter $f_{{\rm NL}}^{{\rm local}}$. Our numerical results show that the predictions of the models for $d{n_s}/d\ln k$ are in agreement with the 95\% CL range of Planck 2015 TT, TE, EE+lowP data \cite{Planck2015}. Also, the evaluated values for $f_{{\rm NL}}^{{\rm local}}$ are compatible with the 68\% CL constrains of Planck 2015 T+E data \cite{Planck2015NG}.

\begin{acknowledgments}

{The authors thank the referees for their valuable comments.}

\end{acknowledgments}










\end{document}